\documentclass[twocolumn,tighten]{aastex63}
\usepackage{amsmath}
\usepackage{threeparttable}
\usepackage{float}
\usepackage{graphicx}
\usepackage{CJK}
\usepackage{xspace}

\newcommand{\package}[1]{\tt{#1}}

\newcommand{\sersic}{S\'{e}rsic \,}

\shorttitle{Schrodinger's Galaxy: $z\approx17$ or $z\approx5$?}
\shortauthors{Naidu \& Oesch et al.}

\begin{document}
\begin{CJK*}{UTF8}{gbsn}

\title{Schrodinger's Galaxy Candidate: Puzzlingly Luminous at $z\approx17$, or Dusty/Quenched at $z\approx5$?
}

\correspondingauthor{Rohan P. Naidu, Pascal A. Oesch}
\email{rohan.naidu@cfa.harvard.edu, pascal.oesch@unige.ch}
\author[0000-0003-3997-5705]{Rohan P. Naidu}
\affiliation{Center for Astrophysics $|$ Harvard \& Smithsonian, 60 Garden Street, Cambridge, MA 02138, USA}
\author[0000-0001-5851-6649]{Pascal A. Oesch}
\affiliation{Department of Astronomy, University of Geneva, Chemin Pegasi 51, 1290 Versoix, Switzerland}
\affiliation{Cosmic Dawn Center (DAWN), Niels Bohr Institute, University of Copenhagen, Jagtvej 128, K\o benhavn N, DK-2200, Denmark}
\author[0000-0003-4075-7393]{David J. Setton}
\affiliation{Department of Physics and Astronomy and PITT PACC, University of Pittsburgh, Pittsburgh, PA 15260, USA}
\author[0000-0003-2871-127X]{Jorryt Matthee}
\affiliation{Department of Physics, ETH Z\"urich, Wolfgang-Pauli-Strasse 27, 8093 Z\"urich, Switzerland}
\author[0000-0002-1590-8551]{Charlie Conroy}
\affiliation{Center for Astrophysics $|$ Harvard \& Smithsonian, 60 Garden Street, Cambridge, MA 02138, USA}
\author[0000-0002-9280-7594]{Benjamin D. Johnson}
\affiliation{Center for Astrophysics $|$ Harvard \& Smithsonian, 60 Garden Street, Cambridge, MA 02138, USA}
\author[0000-0003-1614-196X]{John.~R.~Weaver}
\affiliation{Department of Astronomy, University of Massachusetts, Amherst, MA 01003, USA}
\author[0000-0002-4989-2471]{Rychard J. Bouwens}
\affiliation{Leiden Observatory, Leiden University, NL-2300 RA Leiden, Netherlands}
\author[0000-0003-2680-005X]{Gabriel B. Brammer}
\affiliation{Cosmic Dawn Center (DAWN), Niels Bohr Institute, University of Copenhagen, Jagtvej 128, K\o benhavn N, DK-2200, Denmark}
\author[0000-0001-8460-1564]{Pratika Dayal}
\affiliation{Kapteyn Astronomical Institute, University of Groningen, 9700 AV Groningen, The Netherlands}
\author[0000-0002-8096-2837]{Garth Illingworth}
\affiliation{Department of Astronomy and Astrophysics, University of California, Santa Cruz, CA 95064, USA}
\author[0000-0003-1641-6185]{Laia Barrufet}
\affiliation{Department of Astronomy, University of Geneva, Chemin Pegasi 51, 1290 Versoix, Switzerland}
\author[0000-0002-5615-6018]{Sirio Belli}
\affiliation{Dipartimento di Fisica e Astronomia, Università di Bologna, via Gobetti 93/2, 40122 Bologna, Italy}
\author[0000-0001-5063-8254]{Rachel Bezanson}
\affiliation{Department of Physics and Astronomy and PITT PACC, University of Pittsburgh, Pittsburgh, PA 15260, USA}
\author[0000-0002-0974-5266]{Sownak Bose}
\affiliation{Institute for Computational Cosmology, Department of Physics, Durham University, Durham, DH1 3LE, UK}
\author[0000-0002-9389-7413]{Kasper E. Heintz}
\affiliation{Cosmic Dawn Center (DAWN), Niels Bohr Institute, University of Copenhagen, Jagtvej 128, K\o benhavn N, DK-2200, Denmark}
\author[0000-0001-6755-1315]{Joel Leja}
\affiliation{Department of Astronomy \& Astrophysics, The Pennsylvania
State University, University Park, PA 16802, USA}
\affiliation{Institute for Computational \& Data Sciences, The Pennsylvania State University, University Park, PA, USA}
\affiliation{Institute for Gravitation and the Cosmos, The Pennsylvania State University, University Park, PA 16802, USA}
\author[0000-0002-5757-4334]{Ecaterina Leonova}
\affiliation{GRAPPA, Anton Pannekoek Institute for Astronomy and Institute of High-Energy Physics, University of Amsterdam, Science Park 904, 1098 XH Amsterdam, The Netherlands}
\author[0000-0001-8442-1846]{Rui Marques-Chaves} 
\affiliation{Department of Astronomy, University of Geneva, Chemin Pegasi 51, 1290 Versoix, Switzerland}
\author[0000-0001-7768-5309]{Mauro Stefanon}
\affiliation{Departament d'Astronomia i Astrof\`isica, Universitat de Val\`encia, C. Dr. Moliner 50, E-46100 Burjassot, Val\`encia,  Spain}
\affiliation{Unidad Asociada CSIC "Grupo de Astrof\'isica Extragal\'actica y Cosmolog\'ia" (Instituto de F\'isica de Cantabria - Universitat de Val\`encia)}
\author[0000-0003-3631-7176]{Sune Toft}
\affiliation{Cosmic Dawn Center (DAWN), Niels Bohr Institute, University of Copenhagen, Jagtvej 128, K\o benhavn N, DK-2200, Denmark}
\affiliation{Niels Bohr Institute, University of Copenhagen, Jagtvej 128, K\o benhavn N, DK-2200, Denmark}
\author[0000-0002-5027-0135]{Arjen van der Wel}
\affil{Astronomical Observatory, Ghent University, Krijgslaan 281, Ghent, Belgium}
\author[0000-0002-8282-9888]{Pieter van Dokkum}
\affiliation{Astronomy Department, Yale University, 52 Hillhouse Ave,
New Haven, CT 06511, USA}
\author[0000-0001-8928-4465]{Andrea Weibel}
\affiliation{Department of Astronomy, University of Geneva, Chemin Pegasi 51, 1290 Versoix, Switzerland}
\author[0000-0001-7160-3632]{Katherine E. Whitaker}
\affil{Department of Astronomy, University of Massachusetts, Amherst, MA 01003, USA}
\affiliation{Cosmic Dawn Center (DAWN), Niels Bohr Institute, University of Copenhagen, Jagtvej 128, K\o benhavn N, DK-2200, Denmark}

\begin{abstract}
\textit{JWST}'s first glimpse of the $z>10$ Universe has yielded a surprising abundance of luminous galaxy candidates. Here we present the most extreme of these systems: CEERS-1749. Based on $0.6-5\mu$m photometry, this strikingly luminous ($\approx$26 mag) galaxy appears to lie at $z\approx17$. This would make it an $M_{\rm{UV}}\approx-22$, $M_{\rm{\star}}\approx5\times10^{9}M_{\rm{\odot}}$ system that formed a mere $\sim220$ Myrs after the Big Bang. The implied number density of this galaxy and its analogues challenges virtually every early galaxy evolution model that assumes $\Lambda$CDM cosmology. However, there is strong environmental evidence supporting a secondary redshift solution of $z\approx5$: all three of the galaxy's nearest neighbors at $<2.5\arcsec$ have photometric redshifts of $z\approx5$. Further, we show that CEERS-1749 may lie in a $z\approx5$ protocluster that is $\gtrsim5\times$ overdense compared to the field.  Intense line emission at $z\approx5$ from a quiescent galaxy harboring ionized gas, or from a dusty starburst, may provide satisfactory explanations for CEERS-1749's photometry. The emission lines at $z\approx5$ conspire to boost the $>2\mu$m photometry, producing an apparent blue slope as well as a strong break in the SED. Such a perfectly disguised contaminant is possible only in a narrow redshift window ($\Delta z\lesssim0.1$), implying that the permitted volume for such interlopers may not be a major concern for $z>10$ searches, particularly when medium-bands are deployed. If CEERS-1749 is confirmed to lie at $z\approx5$, it will be the highest-redshift quiescent galaxy, or one of the lowest mass dusty galaxies of the early Universe detected to-date ($A_{5500}\approx1.2$ mag, $M_{\rm{\star}}\approx5\times10^{8}M_{\rm{\odot}}$). Both redshift solutions of this intriguing galaxy hold the potential to challenge existing models of early galaxy evolution, making spectroscopic follow-up of this source critical.

\end{abstract}


\section{Introduction}
\label{sec:introduction}

The first weeks of \textit{JWST} data have already produced an overwhelming number of science publications. At the highest redshifts, \textit{JWST} has finally pushed our observational frontier beyond $z\sim11$, into the last unknown epoch of our cosmic timeline. Several teams have reported a surprisingly large number of particularly luminous sources very early in the Universe's history \citep[e.g.,][]{Naidu22,Castellano22,Donnan22, Atek22,Yan22,Finkelstein22, Harikane22b}.

Some of these early $z>10$ galaxy candidates have proven quite surprising given that the current area surveyed with \textit{JWST}/NIRCam used for these first studies is still rather limited ($<$60 arcmin$^2$). Standard extrapolations of the UV luminosity function's (UV LF) evolution and galaxy simulations predict a much larger survey area is required to yield such a bounty of luminous candidates. Even though the overall evolution of the UV LF at $z>10$ had been debated from the limited information available from \textit{HST} data \citep[e.g.,][]{Oesch18,McLeod16}, evidence had been growing for a differential evolution of the galaxy population: a high number density of the most UV-luminous sources seems to be in place rather early \citep[e.g.,][]{Stefanon19,Bowler20,Morishita20,Harikane22,Bagley22}, while the number density of the fainter galaxy population continues to decline at $z>8$ \citep[e.g.,][]{Oesch18,Ishigaki18,Bouwens22b}. 

Theoretical models based on forming galaxies embedded in dark matter halos have not yet successfully reproduced this differential evolution, should it be confirmed. Most models thus underpredict the number density of the most UV-luminous observed galaxies \citep[e.g.,][]{Bowler20,Leethochawalit22GLASS,Finkelstein22}. Already with \textit{HST}, the discovery of GN-z11 in a very small search volume was very puzzling \citep{Oesch16,Waters16,Mutch16}. Now, the first results from \textit{JWST} imaging are further challenging models. 

However, before changing the theoretical models, it is important to first confirm the high-redshift nature of these galaxies. Since \textit{JWST} provides us with the first-ever deep, high-resolution view of the Universe at $>2$ \micron, it is conceivable that we are discovering other, hitherto unknown galaxy populations. In particular, the most distant Lyman break galaxies (LBGs) are typically selected based on one single spectral feature: the complete break at the redshifted Ly$\alpha$ line due to absorption by the neutral inter-galactic medium. This break can be confused with a Balmer break at lower redshift \citep[e.g.,][]{Vulcani17}, or with extremely strong emission line sources \citep[e.g.,][]{Atek11,vanderWel11}, especially when limited filter coverage is available longward of the break \citep[e.g.,][]{Brammer13}. Additionally, dust-obscured sources can cause very red colors rendering galaxies undetected at shorter wavelengths, even in \textit{JWST} data \citep[e.g.,][]{Barrufet22,Labbe22,Nelson22,Fudamoto22}. However, such sources are usually guarded against in LBG selections by requiring blue SEDs longward of the break. Nevertheless, when selecting high-redshift galaxies in a novel regime, one has to be cautious. 

Here, we present a detailed analysis of an extremely luminous $z\sim17$ galaxy candidate identified in the Early Release Science (ERS) program CEERS \citep[][]{Finkelstein22}. The source was first mentioned in \citet{Naidu22} and presented as a likely $z=16.6$ candidate in \citet{Donnan22} as well as \citet{Harikane22b}. Here, we present a detailed analysis of this source, including arguments for a possible lower redshift solution of this galaxy at $z\sim5$ \citep[see also][]{Zavala22}.

This paper is organized as follows. \S\ref{sec:data} describes the dataset and sample selection. In \S\ref{sec:results} we present our results, followed by a discussion on the implications follows in \S\ref{sec:discussion} and a summary in \S\ref{sec:summary}. 
Magnitudes are in the AB system \citep[e.g.,][]{Oke83}. For summary statistics we report medians along with 16$^{\rm{th}}$ and 84$^{\rm{th}}$ percentiles. We adopt a \citet{Planck2015} cosmology.

\begin{figure*}
\centering
\includegraphics[width=0.9\linewidth]{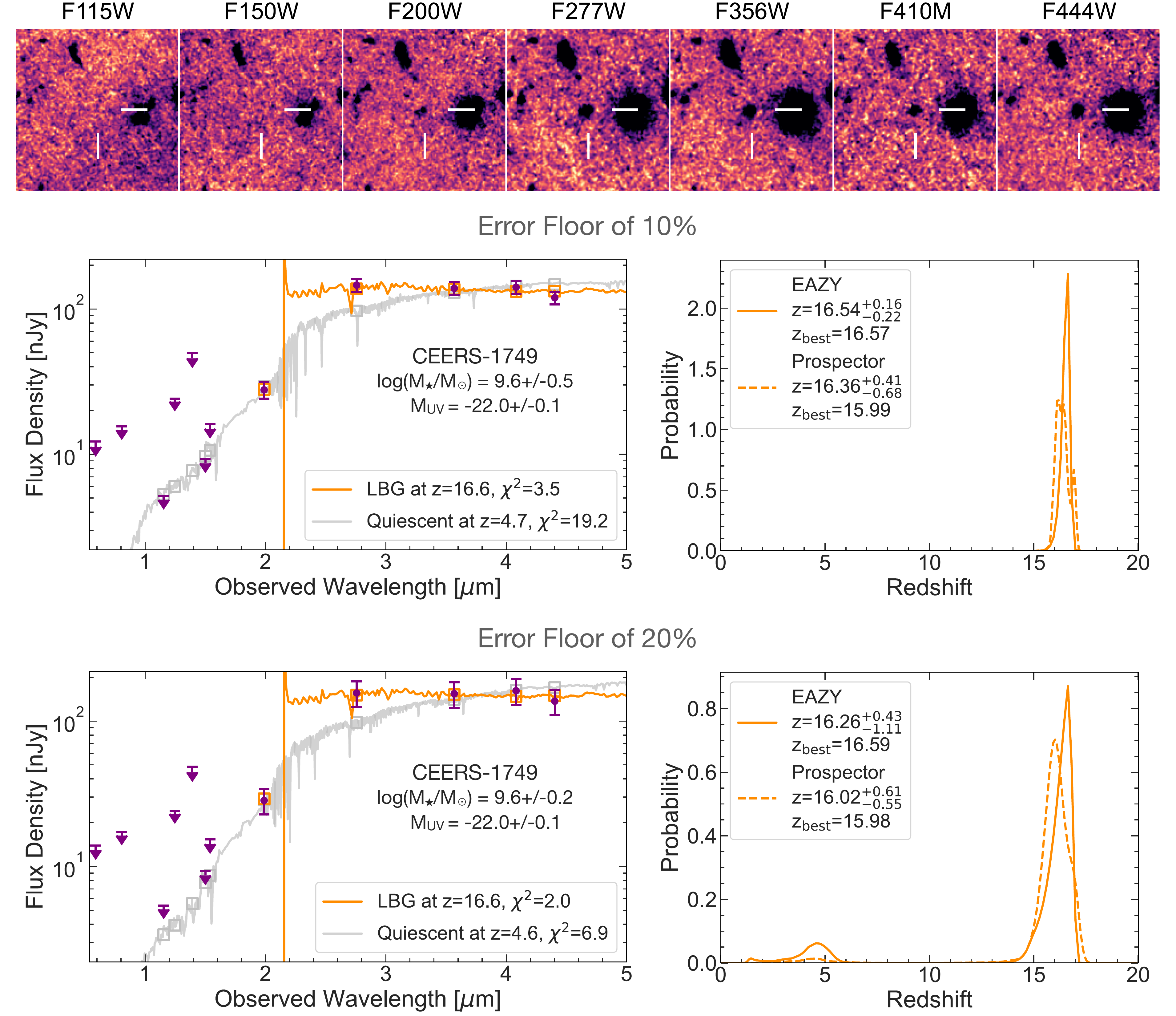}
\caption{Summary of photometry and redshift solutions for CEERS-1749. \textbf{Top:} Multi-wavelength photometry in $3\arcsec$x$3arcsec$ cutouts spanning $\approx1-4.5\mu$m. The source is well-detected ($>9-20\sigma$) in all but the two bluest bands, where it entirely drops out. \textbf{Center:} Redshift solution assuming a $10\%$ error floor on photometry. The data are depicted in purple -- upper limits are shown as arrows, and represent 1-$\sigma$ limits. The best-fit spectral energy distribution (SED) from \texttt{EAZY} is shown in dark orange, representing a Lyman break galaxy (LBG) at $z\approx16.6$. The best-fit SED from our \texttt{EAZY} run constrained to $z<6$ is shown in silver, and corresponds to a quiescent galaxy whose Balmer break produces a drop in flux across F277W and F200W. The failure to match the photometry in F277W coupled with the non-detections in the bluer bands together strongly favor the LBG solution. In the right panel, we show the probability distributions for the redshift derived using \texttt{EAZY} (orange solid) and \texttt{Prospector} (orange dashed). Both frameworks place the source at $z>10$ with $>99.9\%$ probability. \textbf{Bottom:} Same as center panels, but adopting a $20\%$ error floor on photometry. While the best-fit \texttt{EAZY} SEDs in the left-panel are similar as the $10\%$ error floor case, a notable feature in the right-panel is a solution occurring at $z\approx5$ that appears only when this conservative error floor is adopted. See also Figure \ref{fig:altSEDs} for other possible $z\sim5$ SED solutions from \texttt{Prospector}. }
\label{fig:summaryCEERSz17}
\end{figure*}

\section{Data \& Methods}
\label{sec:data}

\begin{deluxetable}{lr}
\label{table:photometry}
\tabletypesize{\footnotesize}
\tablecaption{Photometry in units of nJy -- \textit{JWST}/NIRCam followed by \textit{HST}/WFC3.}
\tablehead{ \colhead{Band} & \colhead{\hspace{2cm} CEERS-1749} }
\startdata
\vspace{-0.2cm}  \\
$F$115$W$ & -1$\pm$5\\
$F$150$W$ & 3$\pm$6\\
$F$200$W$ & 26$\pm$3\\
$F$277$W$ & 127$\pm$5\\
$F$356$W$ & 110$\pm$5\\
$F$410$M$ & 107$\pm$9\\
$F$444$W$ & 91$\pm$5\\
\hline
$F$606$W$ & 5$\pm$6\\
$F$814$W$ & 5$\pm$8\\
$F$125$W$ & 5$\pm$11\\
$F$140$W$ & 9$\pm$24\\
$F$160$W$ & -4$\pm$10\\
\enddata
\tablecomments{We set an error floor of $20\%$ on the fluxes for all \texttt{EAZY} and \texttt{Prospector} fits to conservatively account for systematic uncertainty that is not reflected in the errors here.}
\end{deluxetable}

\subsection{Imaging Data and Catalogs}
The data and analysis methods used here are the same as in \citet[][]{Naidu22}. We refer the reader to that paper for details. Briefly, this work is based on some of the first \textit{JWST}/NIRCam imaging datasets from the Early Release Science (ERS) programs CEERS \citep{Finkelstein22} and GLASS \citep{Treu22}. In particular, the sources analyzed in detail here were identified in the CEERS NIRCam images that span $\sim$40 arcmin$^2$.

The stage 2, calibrated images were obtained from the MAST archive and processed with the \texttt{grizli} pipeline, which performs WCS alignment and image mosaicking. Additionally, \texttt{grizli} masks `snowballs' and mitigates the 1/f noise that are most prominent in the short-wavelength data \citep{Rigby22}.  The \textit{JWST} data used here includes the six wide filters F115W, F150W, F200W, F277W, F356W, F444W and the medium band filter F410M. The 5$\sigma$ depth as measured in empty circular sky apertures of 0\farcs32 diameter ranges from 28.5 to 28.9 mag in the wide filters.

Additionally, we include ancillary \textit{HST} data available in the AEGIS field, most notably from the CANDELS survey \citep{Koekemoer11,Grogin11}. Specifically, we use a re-reduction of the ACS F606W and F814W images, in addition to F125W, F140W, and F160W taken with WFC3/IR. All images were drizzled at 40mas and aligned to the GAIA DR3 catalog.

We use \texttt{SExtractor} to detect sources in the F444W filter and measure multi-wavelength photometry in small circular apertures of 0\farcs32 diameter. These fluxes are then corrected to total using the AUTO flux measurement in our detection band, in addition to applying small corrections for remaining flux lost based on the point-spread functions.

\subsection{NIRCam Zeropoint Uncertainties} 

Given possible uncertainties with the NIRCam zeropoints \citep[see, e.g.,][]{Rigby22}, we perform several tests on the current photometry. In particular, we use the available spectroscopic redshifts for galaxies in the CANDELS/EGS field to derive iterative zeropoint corrections with \texttt{EAZY}. The derived corrections depend on the exact choices of parameters. However, they typically remain smaller than $20\%$. To allow for this systematic uncertainty, we decided to set an error floor of 20\% to all photometric bands, but to use the original pipeline-provided zeropoints. As discussed later, this choice results in the appearance of lower redshift solutions for some de-facto secure very high-redshift candidates.

\subsection{Selection of Luminous High-Redshift Galaxy Candidates} 
This paper follows an earlier analysis of \citet{Naidu22}, who identified the most luminous galaxy candidates in the existing ERS imaging data, based on optical non-detection criteria and photometric redshift measurements using the \texttt{EAZY} code \citep{Brammer08}. As first noted in that paper, the CEERS field revealed a seemingly reliable, but extremely luminous galaxy candidate with photometric redshift at $z=16.6\pm0.1$ \citep[see also][]{Donnan22, Harikane22b}. Given its extraordinary luminosity, if confirmed to lie at $z\sim17$, this source deserved special attention and analysis. We will discuss it in detail below.

\section{Results}
\label{sec:results}

\subsection{Evidence for a $z\approx17$ solution}
\label{sec:candidates}

We briefly discussed CEERS-1749 as a $z\approx17$ candidate in our search for luminous $z>10$ galaxies presented in \citet[][]{Naidu22}. CEERS-1749 is an extremely luminous (26.3 mag in $F$277$W$) galaxy that is well-detected in all bands at $\gtrsim2\mu$m (Figure \ref{fig:summaryCEERSz17}). It shows a sharp drop-off in flux between F277W and F200W (1.7 mag), is undetected at lower wavelengths (F115W, F150W), and appears to display a blue continuum slope characteristic of early Universe galaxies at longer wavelengths. Both \texttt{EAZY} and the \texttt{Prospector} modeling framework \citep[][]{Johnson21} interpret the break in the SED as being due to total absorption of $<1215$\AA\, rest-frame photons at $z\approx17$ by the neutral IGM. The redshift probability distribution is almost entirely contained at $z>10$ -- $p(z>10)>99.9\%$ -- unless a conservative error floor of $20\%$ on the photometry is adopted (see center and bottom panels of Figure \ref{fig:summaryCEERSz17}, discussed below). 

While this candidate satisfied all our quality cuts and search criteria in \citet[][]{Naidu22}, we were concerned that the $F$150$W$ and $F$115$W$ photometry for this source, which are crucial to its candidacy, are based on slightly lower SNR areas of the mosaic, with uneven depth apparent around the object (see e.g., the F115W stamp in Figure \ref{fig:summaryCEERSz17}). This, combined with current NIRCam calibration uncertainties (\citealt{Rigby22}) made judging a stringent flux upper limit difficult. Stringent non-detections in these bands are critical to the source's candidacy because the break in the SED by itself is unable to settle the case. Unlike in the cases of GLASS-z11 and GLASS-z13 \citep[][]{Naidu22,Castellano22}, the drop in flux across immediately adjacent filters is not as dramatic ($\approx5\times$ vs. $\approx10\times$), leaving more room for other possibilities (as we discuss below).

We have since performed a battery of tests to check the robustness of the $z\approx17$ solution. The most crucial of these is the incorporation of \textit{HST} data, in particular, deep imaging in the F160W band (whose zero-point is well-known) where the source is undetected with a stringent $<10$ nJy $1\sigma$ upper-limit, supporting the NIRCam/F150W non-detection that we report here. In fact, the \texttt{EAZY} redshift derived by replacing the bluest \textit{JWST} bands (F115W and F150W) with \textit{HST}/F160W yields $z=16.5$ as the most likely solution. Other notable tests include recovering this source at $z\approx17$ in an entirely independent analysis of the CEERS images (Bouwens et al., in prep.), and setting a conservative error-floor of $\approx20\%$ on \textit{all} fluxes to account for systematic uncertainty in e.g., zeropoints across this analysis.

If confirmed at $z\approx17$, CEERS-1749 would be one of the most luminous galaxy candidates at $z>10$ ($M_{\rm{UV}}\approx-22$), second only to HD1 ($M_{\rm{UV}}\approx-23$, \citealt{Harikane22,Pacucci22}) and comparable to the spectroscopically confirmed GNz11 \citep[][]{Oesch16,Jiang21}. A source of such luminosity simply does not exist at such a potentially early time in a wide swath of empirical and theoretical models of the $z>10$ Universe (discussed further in \S\ref{sec:discussionz17}).

\subsection{Evidence for a $z\approx5$ solution}
\label{sec:lowz}

While the $z\approx17$ solution is formally favored, there is non-zero probability that the source lies at $z\approx5$.   We note that our $p(z)$ estimate does not include a prior based on the luminosity function (which is poorly constrained at these epochs and luminosities).  So, while the low redshift solution is formally disfavored, it is not ruled out, and the relative probability of the two solutions depends sensitively on our prior belief of the source residing at $z\approx5$ compared to $z\approx17$. In this section we consider the evidence for the lower redshift solution.

\subsubsection{Environmental evidence: a $z\approx5$ protocluster?}

\begin{figure*}
\centering
\includegraphics[width=0.95\linewidth]{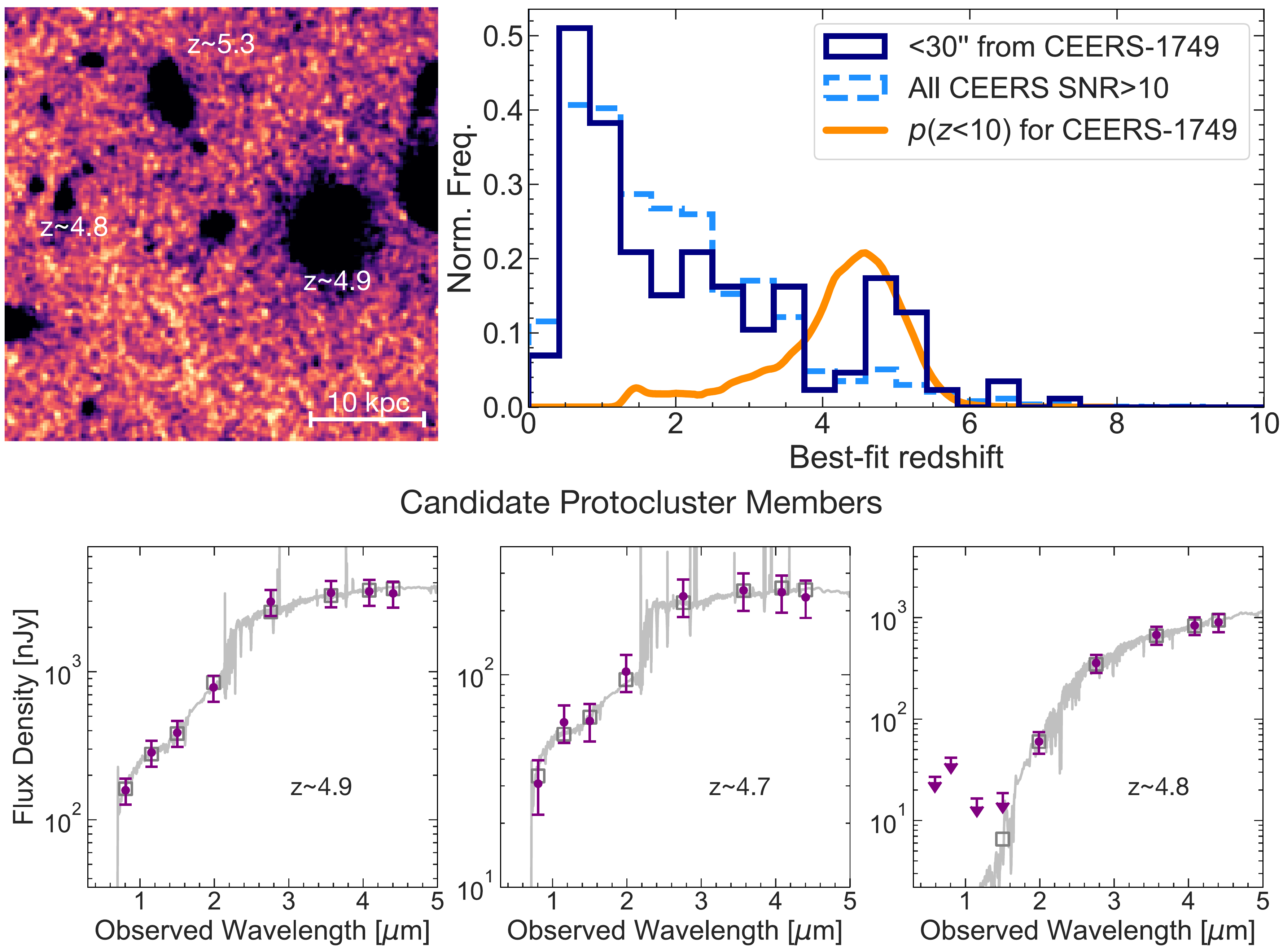}
\caption{Environmental support for a $z\approx5$ solution for CEERS-1749. \textbf{Top Left:} $3.4\arcsec$x$3.4\arcsec$ F277W image centered on CEERS-1749. The best-fit photometric redshifts of the three nearest neighbors are labeled in white. All neighbors lie at $z\approx5$, which happens to be the predicted redshift for CEERS-1749 if it were a quiescent galaxy. \textbf{Top Right:} Best-fit redshift distribution in 30'' around CEERS-1749 (solid navy blue line) compared to the redshift distribution of the whole CEERS field (dashed light blue line) for galaxies detected at $>10\sigma$ in all LW bands. The redshift probability distribution for CEERS-1749 at $z<10$ is overlayed in orange. The redshift distribution in the immediate vicinity of CEERS-1749 shows a pronounced excess at $z\approx5$, with $\approx20$ galaxies at $4.5<z<5.5$. That is, not only are CEERS-1749's immediate neighbors at $z\approx5$, but it also lies amidst a $z\approx5$ protocluster candidate. \textbf{Bottom:} Best-fit \texttt{EAZY} SEDs of some of the potential protocluster members show clear signatures of ancient stellar populations (e.g., strong Balmer breaks) consistent with the early formation epoch expected of such clusters, making it more plausible that we are indeed observing a protocluster.}
\label{fig:z5solution}
\end{figure*}

The first argument we consider for the $z\approx5$ solution is based on the local environment and large-scale structure near the source.  The galaxy's three nearest neighbors \textit{all} lie at precisely the redshift required by the quiescent galaxy solution, i.e., $z\approx5$ (Figure \ref{fig:z5solution}, top-left). The most massive of these galaxies is the nearest neighbor only $1.7\arcsec$ away at $z_{\rm{EAZY}}=4.9$ whose mass we fit to be a substantial $M_{\rm{\star}}\approx10^{11} M_{\rm{\odot}}$ (via the fiducial \texttt{Prospector} setup described in \citealt[][]{Naidu22}). If CEERS-1749 lies at the same redshift, and therefore a physical separation of $<15$ kpc, it may be an associated satellite galaxy of its massive neighbor analogous to the Magellanic Clouds ($M_{\rm{\star}}\approx5\times10^{9} M_{\rm{\odot}}$, \citealt[][]{vandermarel09}) and the present-day Milky Way ($M_{\rm{\star}}\approx10^{11} M_{\rm{\odot}}$, \citealt[][]{Bland-Hawthorn16}).

CEERS-1749 may also be part of a larger $z\approx5$ protocluster. We find that the $30\arcsec$ region around CEERS-1749 is overdense in galaxies with best-fit photometric redshifts of $z\approx5$ compared to the overall CEERS field (Figure \ref{fig:z5solution}). Across the full 40 arcmin$^{2}$ analyzed we find $\approx300$ galaxies at $4.5<z<5.5$ with SNR$>10$ in all LW bands (F444W, F356W, F277W). Strikingly, in $30\arcsec$ around CEERS-1749 we find $\approx20$ such sources, which translates to a $\approx4\times$ higher density than the field average. Even taking a model-independent point of view, selecting sources akin to CEERS-1749 with a red F200W-F277W color ($>0.75$) that are detected at $<2\sigma$ in F606W (the dropout filter at $z\approx5$) shows an overdensity of $\approx9\times$ in the $30\arcsec$ around the source. Intriguingly, the median redshift of these potential protocluster galaxies ($z_{\rm{cluster}}=4.9$) is an excellent match to the predicted range for a lower-redshift solution. 

In models of hierarchical structure formation, protoclusters are expected to be among the first sites of star-formation in the Universe, as they form from the first overdensities that collapse into stars. Intriguingly, in the $z\approx5$ sample proximal to CEERS-1749 we do find galaxies with strong Balmer breaks characteristic of old stellar populations (see bottom row of Figure \ref{fig:z5solution}). A handful of these galaxies have an F200W-F277W color almost as red as CEERS-1749 within errors, and a handful even show a second mode in their redshift probability distributions at $z\approx16$. The existence of such proximal ancient galaxies at the right redshift makes it more plausible that CEERS-1749 may also belong to this protocluster. 

\subsubsection{Plausible SEDs at $z\approx5$}
\label{sec:whatisz5}

\begin{figure}
\centering
\includegraphics[width=\linewidth]{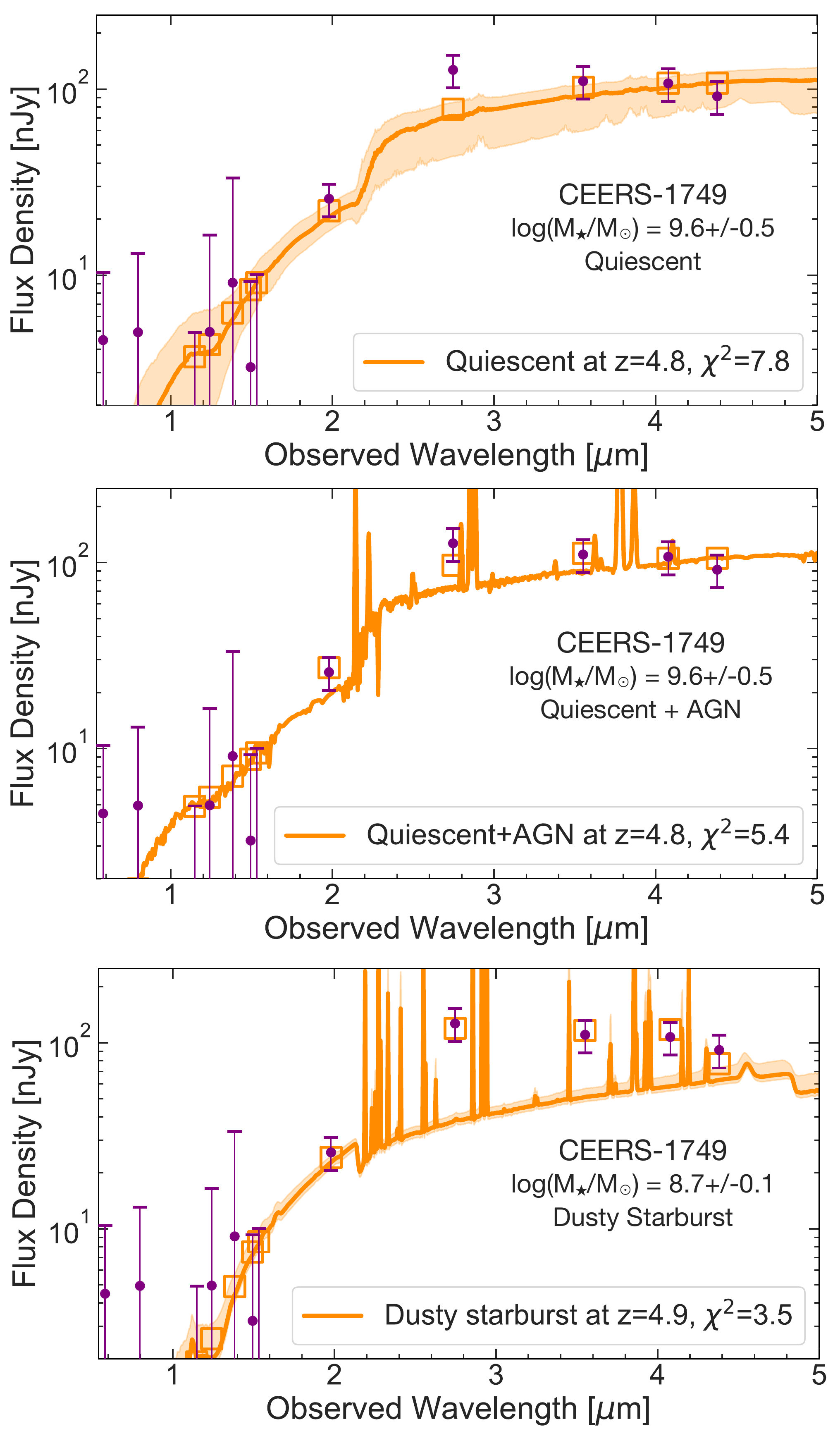}
\caption{\texttt{Prospector} posterior SEDs for CEERS-1749's three classes of $z\approx5$ solutions found by searching around the overdensity redshift ($z=4.9\pm0.1$). The median spectrum is shown in dark orange, with the region between the $16^{\rm{th}}$ and $84^{\rm{th}}$ percentiles shaded. $\chi^{2}$ values are computed assuming a $20\%$ error floor on photometry. \textbf{Top:} Quiescent galaxy solutions where the Balmer break straddles F200W and F277W. The F277W prediction falls short of the observed flux by $\approx50\%$. \textbf{Center:} The highest likelihood quiescent solution, now with emission arising from ionized gas around an AGN. This type of emission helps explain the F200W-F277W color. While the required AGN luminosity is an outlier as per local scaling relations, any mechanism that ionizes gas (e.g., shock heating in the protocluster environment) could produce such an SED. \textbf{Bottom:} A dusty starburst with intense nebular emission. All the $>2\mu$m bands, particularly F277W containing [OIII]$\lambda5007,4959$\AA+H$\beta$, are boosted by line emission, while flux in the bluest bands is suppressed by a steep attenuation curve that mimics a $z>10$ dropout.}
\label{fig:altSEDs}
\end{figure}

\begin{deluxetable}{lrrrrrrrrrrrrr}
\label{table:miri}
\tabletypesize{\footnotesize}
\tablecaption{Predicted \textit{JWST}/MIRI photometry in units of nJy shows significant differences at longer wavelengths across the scenarios discussed in \S\ref{sec:whatisz5}.}
\tablehead{
\colhead{Band} & \colhead{$z\approx17$} & \colhead{$z\approx5$} & \colhead{$z\approx5$}\\
\colhead{} & \colhead{} & \colhead{Quiescent} & \colhead{Starburst}}
\startdata
\vspace{-0.2cm}  \\
F560W & 99$^{+11}_{-10}$ & 133$^{+19}_{-21}$ & 109$^{+22}_{-7}$\\ 
F770W & 109$^{+55}_{-20}$ & 136$^{+26}_{-24}$ & 83$^{+28}_{-5}$\\ 
F1000W & 87$^{+68}_{-29}$ & 136$^{+37}_{-32}$ & 84$^{+37}_{-5}$\\ 
F1280W & 106$^{+93}_{-39}$ & 118$^{+37}_{-27}$ & 169$^{+21}_{-11}$\\ 
F1500W & 71$^{+102}_{-27}$ & 86$^{+36}_{-21}$ & 91$^{+22}_{-7}$\\ 
F1800W & 78$^{+107}_{-26}$ & 82$^{+82}_{-23}$ & 199$^{+49}_{-16}$\\ 
F2100W & 77$^{+96}_{-37}$ & 69$^{+81}_{-21}$ & 228$^{+49}_{-17}$
\enddata
\end{deluxetable}

Taking into account that CEERS-1749 may be a part of a $z\approx4.9$ protocluster, we can refine our models for the source by fixing the redshift to that of the overdensity (a Gaussian with $z=4.9\pm0.1$), which significantly shrinks the parameter space search volume.

Under our fiducial $\approx20\%$ error floor assumption, the galaxies we find are all quiescent systems (see top-row of Figure \ref{fig:altSEDs}). The only data point that this class of solutions struggles to explain is the F277W flux, where the prediction falls short by $\approx50\%$ -- this is why these solutions are disfavored compared to the $z\approx17$ scenario in e.g., Figure \ref{fig:summaryCEERSz17}. Perhaps there is a calibration or zero-point issue in our F277W photometry (though \citealt{Donnan22} report a similar F200W-F277W color as in this work). Or perhaps, the models we have employed here are missing some physics (e.g., line emission from AGN or shocks) that manifests around rest-frame 4300-5300 \AA. 

Inspired by this, we add an AGN emission line template (following line ratios from \citealt[][]{Richardson14}, their Table 3) to the quiescent galaxy solution\footnote{https://github.com/bd-j/prospector/blob/agnlines/}. We are also motivated by the finding that ionized gas, likely due to AGN, is routinely observed in quiescent and post-starburst systems \citep[e.g.,][]{Belli19}. The resulting emission lines boost the F277W flux, providing a better fit (center panel, Figure \ref{fig:altSEDs}). However, the required AGN luminosity  -- (H$\beta/L_{\rm{\odot}})/(M_{\rm{\star}}/M_{\rm{\odot}}) \approx10^{-2.5}$ -- is $\gtrsim100\times$ higher than expected for the galaxy's stellar mass based on $z\lesssim2$ scaling relations and observations \citep[e.g.,][]{Heckman14}. Nonetheless, the key takeaway from this exercise is that ionized gas regardless of its origin (e.g., from shock heating in the overdense environment) that produces emission line luminosities and ratios atypical for star-forming galaxies, and is therefore missed by standard models, may help explain the SED.

Finally, we perform a search by adopting a $10\%$ error-floor on all photometry. This yields a starburst solution (bottom panel, Figure \ref{fig:altSEDs}) in a relatively low-mass ($\approx5\times10^{8} M_{\rm{\odot}}$) system. The SED is dominated by young stars, and shows a Balmer \textit{jump} instead of a Balmer break. The $>2\mu$m photometry is remarkably well-explained by nebular emission -- this is a challenging constraint to match, given the required blue slope and the sensitivity of the F410M medium-band. In bluer bands the predicted flux lies barely under the $\approx1\sigma$ detection limits due to significant dust attenuation ($A_{\rm{5500}}\approx1.2$ mag). It is perplexing as to why this solution, with a $\chi^{2}$ lower than the quiescent galaxies, was missed in our broad $z=0-20$ search despite using conservative \texttt{dynesty} sampling settings. A clue might be the extremely tight redshift posterior of $z=4.87^{+0.00}_{-0.02}$. This is consistent with a highly ``spiky" likelihood space, where there is a very precise combination of redshift and nebular parameters that produce the perfect conspiracy of emission lines to mimic the $z\approx17$ system. Constraining the redshift to the overdensity mean, thereby greatly reducing the prior volume, likely helped reveal this solution, but this is a critical issue for future study.

\subsection{Physical Properties}
\label{sec:physical}

We derive physical properties for the source using \texttt{Prospector} \citep[][]{Leja17,Johnson21}. The priors and parameter choices are as described in \citet[][]{Naidu22} (their \S4.1; closely following \citealt{Tacchella22}). Briefly, we fit a non-parametric star-formation history assuming a prior on the redshift that follows the \texttt{EAZY} photometric redshift constraint. For our fiducial run we assume a ``continuity" prior on the star-formation history and a formation redshift of $z=20$ that produces smooth histories disfavoring abrupt jumps from bin to bin \citep[][]{Leja19}. The resulting properties are summarized in Table \ref{table:properties}. In what follows we split the discussion assuming the $z\approx17$ and $z\approx5$ solutions.

\subsubsection{CEERS-1749 at $z\sim17$}

Broadly, CEERS-1749 at $z\sim17$ has properties characteristic of galaxies found at $z\gtrsim8$ \citep[e.g.,][]{Stefanon21,Tacchella22,Leethochawalit22GLASS}, with a star-formation rate expected of its stellar mass and a blue $\beta_{\rm{UV}}\approx-2.3$ \citep[see also \texttt{Bagpipes} fits in][]{Donnan22}. As noted earlier, its $M_{\rm{UV}}\approx-22$ would place it among the most UV-luminous sources at $z\gtrsim6.5$. Perhaps the most striking property from the SED fitting is the stellar mass -- we infer $\log({M_{\rm{\star}}/M_{\rm{\odot}}})\approx9.6$ in stars to have formed in a mere $\sim220$ Myrs after the Big Bang. We caution that estimates of the stellar mass from rest-UV photometry alone come with significant systematic uncertainties since the light is dominated by young stars.

To understand the range of allowed stellar masses we consider the following changes to our fiducial fitting setup -- we assume a ``bursty" prior for the star-formation history that allows for rapid fluctuations \citep[][]{Tacchella22}; we consider three different initialization points for the star-formation history ($z=18, 20, 30$); we consider a Kroupa IMF in addition to the fiducial \citet[][]{Chabrier03} IMF. Across all these various parameter choices, we recover stellar masses of $\approx10^{9.4}-10^{9.8} M_{\rm{\odot}}$ with $\lesssim$0.3 dex uncertainties on each individual fit. As a limiting case we also model the galaxy as a single stellar population set to an age $<200$ Myr (the age of the Universe at $z=16.6$ is 230 Myrs), which yields a stellar mass of  $\log(M_{\rm{\star}}/M_{\rm{\odot}})\approx10.0^{+0.2}_{-0.2}$ and an age of $70^{+30}_{-10}$ Myrs. Based on these experiments, we conservatively adopt a mass of $\log(M_{\rm{\star}}/M_{\rm{\odot}})\approx9.6^{+0.5}_{-0.5}$ or $\approx 5\times10^{9} M_{\rm{\odot}}$ for this source for the discussion in the rest of the paper.

CEERS-1749 was also recently reported by \citet[][see also \citealt{Harikane22b}]{Donnan22}, in their search for $z>10$ candidates across the \textit{JWST} ERS data, as lying at $z=16.6$. They infer a slightly lower mass of $\log(M_{\rm{\star}}/M_{\rm{\odot}})=9.0\pm0.4$. On top of differences between inference frameworks -- \texttt{Bagpipes} vs. \texttt{Prospector}, and importantly their underlying stellar isochrones, \texttt{Parsec} vs. \texttt{MIST},  see e.g., \citealt{Whitler22} -- their reported fluxes are significantly lower than those reported here (by $\approx2\times$), which can likely be traced to aperture corrections. CEERS-1749 is an extended source, and the point-source correction applied for its flux in \citet[][]{Donnan22} may have been insufficient.

\subsubsection{CEERS-1749 at $z\sim5$}

We run fits for CEERS-1749 by fixing the redshift to the protocluster redshift ($z=4.9\pm0.1$) as described in the previous section. We first discuss the quiescent case followed by the low-mass starburst scenario.

The quiescent galaxy solution displays very modest ongoing star-formation -- averaged over the last 50 Myrs, its SFR is $0.1^{+3.6}_{-0.1} M_{\rm{\odot}}$/yr, and over the last 10 Myrs an even more humble $0.02^{+4.08}_{-0.02}M_{\rm{\odot}}$/yr. The constraint against recent star-formation is a direct result of having to produce a strong Balmer break to match the red F200W-F277W color. We estimate its stellar mass to be $\approx5\times10^{9} M_{\rm{\odot}}$. No quiescent galaxy with such a low stellar mass has been spectroscopically confirmed at $z\gtrsim2$ yet, but this has been largely due to the impracticality of continuum spectroscopy at such luminosities. The low-mass quenched population certainly exists, as evidenced by \citet[][]{Marchesini22}'s recent confirmation of a $\approx8\times10^{9}M_{\rm{\odot}}$ quiescent system at $z\approx2.4$ via \textit{JWST}/NIRISS spectroscopy. Other candidate quiescent systems at $z>2$ of even lower mass have been observed as part of the same program, and may be confirmed as NIRISS calibrations improve.

The dusty star-forming galaxy scenario features a system that has formed $\approx15\%$ of its total stellar mass in only the last $<10$ Myrs. Its young stars are producing intense nebular emission that strongly boosts the broadband photometry and comfortably accounts for the F200W-F277W color. Such extreme emission has been inferred to be routine \citep[e.g.,][]{debarros19}, and is now beginning to be directly observed during these epochs \citep[e.g.,][]{Schaerer22}. The high attenuation ($A_{\rm{5500}}\approx1.2$ mag), on the other hand, may seem unexpected for a galaxy of such low stellar mass based on $<2\mu$m-selected samples \citep[e.g.,][]{Cullen18}. However, as fainter members of the ``\textit{HST}-dark" population (galaxies detected purely at $>2\mu$m) come into view \citep[e.g.,][]{Barrufet22,Nelson22,Fudamoto22}, galaxies like CEERS-1749 may be found to be common. In fact, the colors of the overdensity suggest it may be the perfect place to prospect for the low-mass end of this population. The dust-curve we infer is slightly steeper than that of the Small Magellanic Cloud ($A_{\rm1500}/A_{\rm{5500}}\approx6$) -- such steep curves are often observed in the local Universe \citep[][]{Salim20}, but typically in lower $A_{\rm{5500}}$ systems \citep[][]{Chevallard13}.

We end this section by observing that broadly, the properties of CEERS-1749 are physically plausible, but fall firmly in observational regimes that are only beginning to come into view with \textit{JWST} (e.g., low-mass quiescent, low-mass \textit{HST}-dark). Their novelty represents an exciting opportunity for surprises, as well as a potential challenge for $z>10$ searches.

\begin{deluxetable}{lrrrrrrrrrr}
\label{table:properties}
\tabletypesize{\footnotesize}
\tablecaption{Summary of properties for the high- and lower-redshift solutions.}
\tablehead{
\colhead{CEERS-1749\tablenotemark{$\dagger$} at} & \colhead{$z\sim17$} & \colhead{$z\sim5$} & \colhead{$z\sim5$}\\
\colhead{} & \colhead{} & \colhead{(quiescent)} & \colhead{(starburst)}
}
\startdata
\vspace{-0.2cm}  \\
R.A. & \multicolumn{3}{c}{14:19:39.48} \\
Dec. & \multicolumn{3}{c}{+52:56:34.95} \\
Best-fit Redshift $z_{\rm{Prospector}}$ & $16.0$ & $4.8$ & $4.9$\\
Redshift $z_{\rm{Prospector}}$ & $16.0^{+0.6}_{-0.6}$ & $4.8^{+0.1}_{-0.1}$ & $4.87^{+0.00}_{-0.02}$\\
Best-fit Redshift $z_{\rm{EAZY}}$ & $16.6$ & $4.6$ & --\\
Redshift $z_{\rm{EAZY}}$ & $16.3^{+0.4}_{-1.1}$ & $4.4^{+0.6}_{-1.0}$ & --\\
UV Luminosity ($M_{\rm{UV}}$) & $-22.0^{+0.1}_{-0.1}$ & $-15.6^{+1.1}_{-0.7}$ & $-14.3^{+0.2}_{-1.0}$ \\
UV Slope ($\beta$; $f_{\rm{\lambda}}\propto \lambda^{\beta}$) & $-2.3^{+0.1}_{-0.1}$ & $1.1^{+1.1}_{-1.2}$ & $3.0^{+0.3}_{-0.5}$\\
Stellar Mass $\log$($M_{\rm{\star}}/M_{\rm{\odot}}$) & $9.6^{+0.2}_{-0.2}$ & $9.6^{+0.2}_{-0.5}$ & $8.7^{+0.1}_{-0.1}$\\
Age ($t_{\rm{50}}$/Myr) & $54^{+27}_{-27}$ & $564^{+105}_{-492}$ & $580^{+17}_{-236}$\\
SFR$_{\rm{50\ Myr}}$ ($M_{\rm{\odot}}$/yr) &  $34^{+17}_{-12}$ & $0.1^{+3.6}_{-0.1}$ & $1.7^{+1.8}_{-0.2}$\\
$r_{\mathrm{eff}}$, $F$444$W$ [kpc] & 0.4 & \multicolumn{2}{c}{0.9}\\ 
$r_{\mathrm{eff}}$, $F$200$W$ [kpc] & 0.2, 0.3 & \multicolumn{2}{c}{0.5, 0.7}\\
\enddata
\tablenotetext{\dagger}{This is the same source as CEERS-93316 in \citet{Donnan22} and CR2-z17-1 in \citealt[][]{Harikane22b}.}
\tablecomments{Quantities derived via SED fitting assume a continuity prior (\citealt{Leja19}) on the star-formation history and a \citet[][]{Chabrier03} IMF. Variations on this fiducial assumption, with a focus on the recovered stellar mass, are tested in detail in \S\ref{sec:physical}. The two effective radii quoted in $F$200$W$ are for the two component fit in Fig. \ref{fig:sizefits}.}
\end{deluxetable}

\begin{figure*}
\centering
\includegraphics[width=0.95\linewidth]{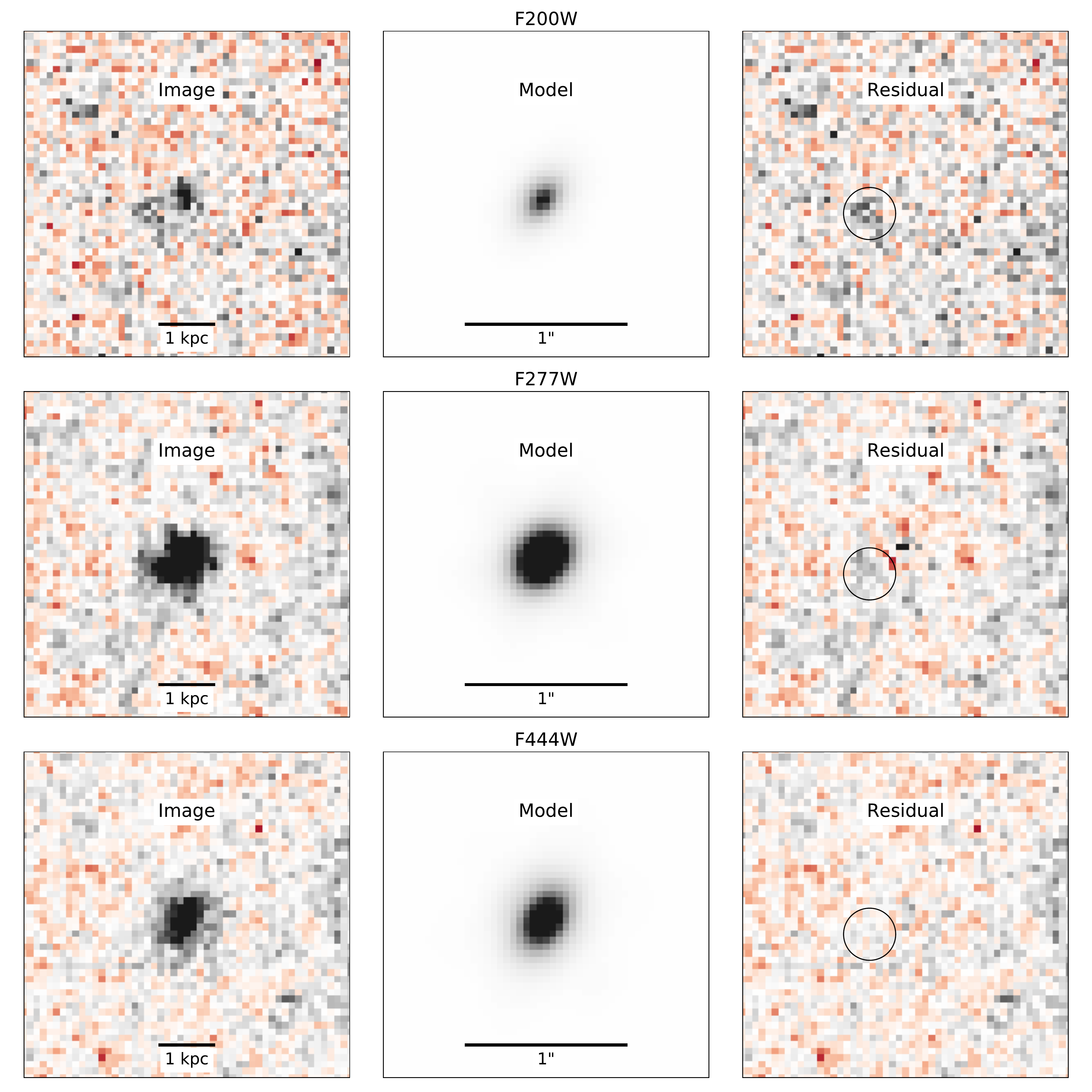}
\caption{Results of the \texttt{GALFIT} morphology analysis for CEERS-1749. The different columns from left to right correspond to the original data, the model, and the residual, and the rows correspond to the analysis of the image in F200W, F277W, and F444W respectively. The images are scaled so that color corresponds to negative flux, black corresponds to postive flux, and white corresponds to 0 flux. The galaxy is consistent in all band with a disky ($n\sim1$) \sersic profile. In the bluer bands, the structure of the galaxy is asymmetric, and a clumpy feature can be seen to the SE (illustrated with a black circle in the residuals). However, in the reddest band, F444W, this feature is absent.}
\label{fig:sizefits}
\end{figure*}

\subsection{Clues from Morphology}
\label{sec:sizefits}

In order to characterize the morphology of the galaxy while accounting for the effect of the PSF, we fit two-dimensional \sersic profiles to the candidate in the F200W, F277W, and F444W imaging using \texttt{GALFIT} \citep{Peng10Galfit}. We create 150-pixel cutouts around the galaxy, then use \texttt{photutils} and \texttt{astropy} to create a segmentation map to identify nearby galaxies using the F444W image. We mask all nearby galaxies using the resultant segmentation maps and fit only CEERS-1749. In all our fits, we measure a scalar sky background estimated from the median of the sky pixels identified in the segmentation map and fix the flux of the sky to this value in \texttt{GALFIT}. We use a theoretical PSF model generated from WebbPSF at our 0\farcs04 pixel scale; we oversample the PSF by a factor of 9 in order to minimize artifacts as we rotate the PSF to the CEERS observation angle calculated from the APT file, then convolve with a 9x9 pixel square kernel and downsample to the mosaic resolution. In all fits, we constrain \sersic index $n$ to be between 0.01 and 8, the magnitude to be between 0 and 45 mag, and the half-light radius $r_e$ to be between 0.3 and 200 pixels (0\farcs012 - 8\farcs0).

The results of these fits are shown in Figure \ref{fig:sizefits}. In all 3 bands, the galaxy is well described by a disky ($n\sim1$) \sersic profile. However, in the F200W image and resultant residuals, there is a clear indication of non-\sersic flux to the southeast of the main galaxy, illustrating that the galaxy may be a clumpy disk or a merging pair. This same residual structure is marginally visible in the F277W image, but in F444W, the galaxy profile is consistent with being completely smooth.

Assuming that CEERS-1749 is at $z\approx17$, its morphology is consistent with bright galaxies catalogued across the Epoch of Reionization, which often break up into clumpy substructure in their rest-UV \citep[e.g.,][]{Bowler17, Matthee19}. The size in $F$444$W$ (0.4 kpc) as well as that of the individual components in $F$277$W$ (0.2 and 0.3 kpc) is quite compact, similar to that measured recently in the CEERS field for the most massive star-forming galaxies at $z\approx7-10$ \citep[][]{Labbe22}, hinting at an evolutionary link between the most luminous $z>10$ systems and the most massive galaxies at lower redshifts.

For the case of CEERS-1749 at $z\approx5$, the nearby clump in the images appears across a wide-range of rest-optical wavelengths ($\approx3000-7500$\AA). Clumpy star-formation occurring in the same disk is a possibility. However, these clumps are typically evident in the UV, with optical profiles behaving more smoothly. 

We might perhaps be witnessing a merger. If CEERS-1749 lives in a dense protocluster environment at $\approx5$, mergers are quite likely. Mergers are a key channel for setting off the kind of vigorous starbursts required for the dusty starburst scenario \citep[e.g.,][]{Mihos96}. Further, several quenching mechanisms, particularly at $z\gtrsim2$, invoke mergers  \citep[e.g.,][]{Man18} -- e.g., an AGN is triggered by fresh inflows of gas to the centre of the galaxy, and the subsequent feedback from the freshly fed AGN quenches star-formation. It is interesting to note that the merging clump in the images appears only in the bluer bands, and is not apparent in F444W. In the $z\approx17$ scenario, strong wavelength-dependent morphology across such a narrow rest-$1600$\AA\, to rest-$2150$\AA\, range is surprising, but is perhaps more easily explained by a blue galaxy merging with a quiescent/dusty one at $z\approx5$.

Another possibility is that CEERS-1749 may be in the process of tidal disruption by its massive neighbor if they lie at the same redshift. The galaxy lies well within the expected virial radius of its neighboring $M_{\rm{\star}}\approx10^{11} M_{\rm{\odot}}$ system that is only $<15$ kpc away. The faint substructure in the images may be stripped tidal debris trailing the galaxy. Starbursts often occur in the disrupting galaxy \citep[e.g.,][]{dicintio21}, following which it loses all its gas and is quenched \citep[e.g.,][]{Teppergarcia18}. A circular polar orbit that decays gradually may allow for a significant period between infall and tidal dissolution that may allow us to observe the galaxy in this transient state (see e.g., the gallery of mergers in \citealt{Naidu21}).

\section{Discussion}
\label{sec:discussion}

\begin{figure*}
\centering
\includegraphics[width=0.48\linewidth]{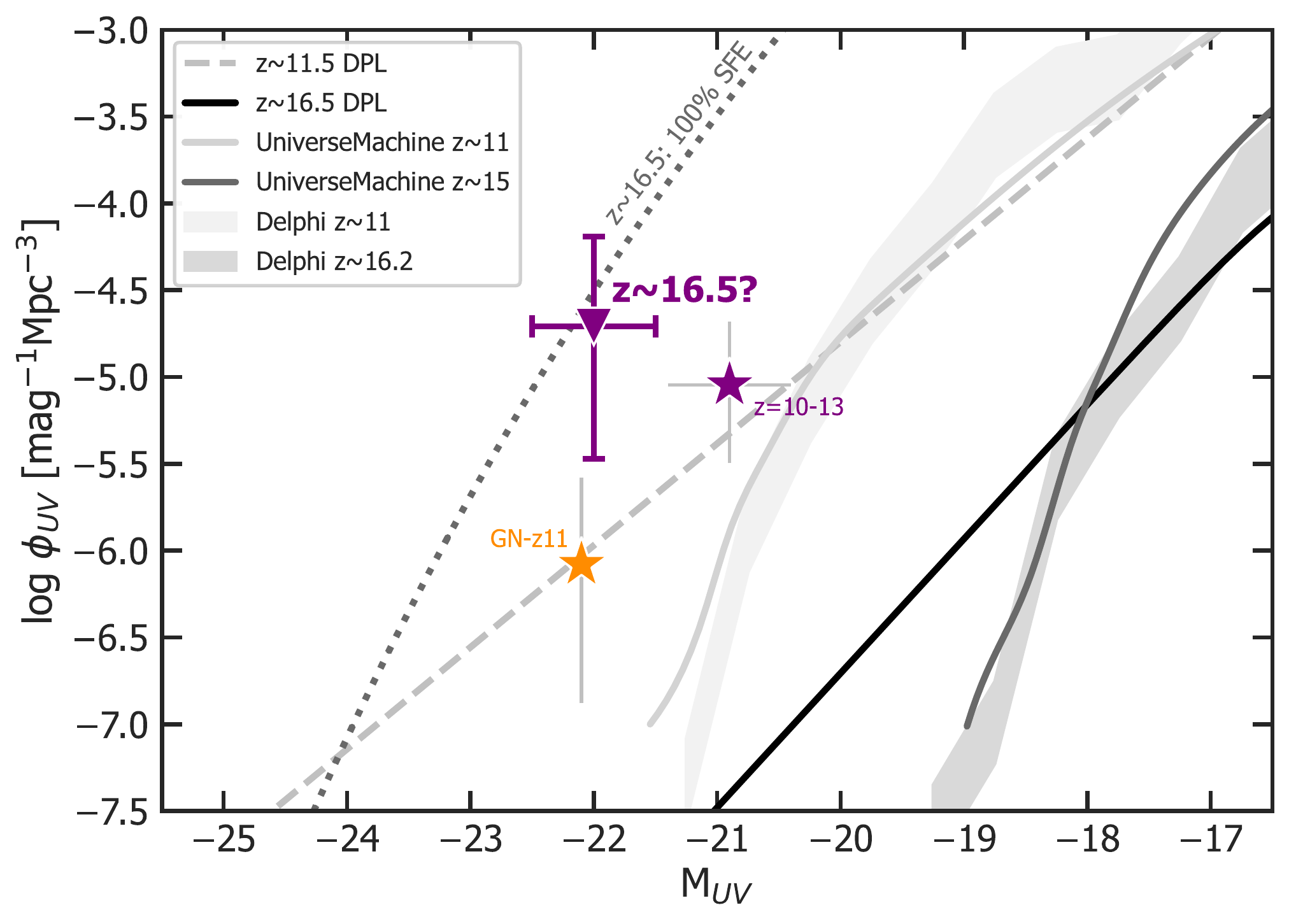} \hspace{0.4cm}
\includegraphics[width=0.46\linewidth]{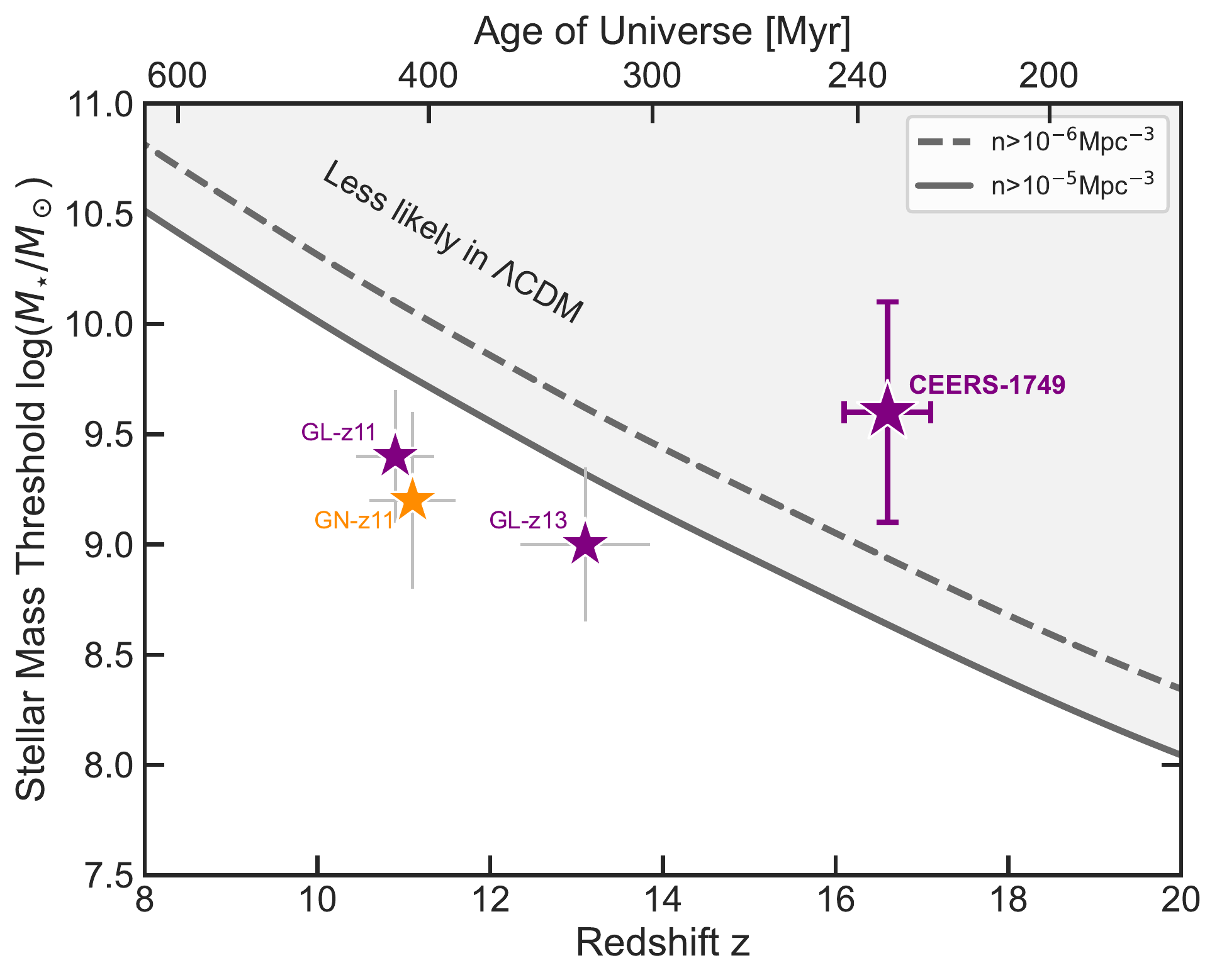}
\caption{How CEERS-1749 at $z\sim16.5$ would challenge our understanding of galaxy formation. \textbf{Left}: Implied constraints on the bright end of the UV LF at $z\sim16.5$. If this source is indeed confirmed to lie at $z\sim16.5$, it would defy virtually every model of early galaxy evolution. The solid light gray lines show predictions from two models \citep[][]{Dayal14,Behroozi19} at $z\sim11$, while the darker gray lines are for $z\sim15-16$. At the observed number density, CEERS-1749 would be 4mag too bright for the model predictions. Even more impressive is the comparison with the DM halo mass function. The dotted gray line shows the predicted LF of an extreme model where \textit{all} baryons in a given halo are converted into stars. This is the only way to reproduce the UV LF of this galaxy. For context, similarly luminous galaxies at $z\approx6-10$ have star-formation efficiencies inferred to be $<10\%$ \citep[e.g.,][]{Tacchella18,Stefanon21mass}. As an empirical comparison we also show extrapolations of the double-power law UVLF (''DPL") from \citet[][]{Bowler20} to $z\approx11$ and $z\approx16$.
\textbf{Right}: Galaxy stellar mass threshold vs. redshift expected for $\Lambda$CDM cosmology, adopted from \citet[][]{Behroozi18}. The stellar mass threshold is derived from halo mass functions that assume a $100\%$ star-formation efficiency, i.e., $M_{\rm{\star}}/M_{\rm{halo}}=f_{\rm{baryon}}$, where $f_{\rm{baryon}}=0.16$ is the cosmic baryon fraction. CEERS-1749, and the tentative implied number density of its analogues ($\approx10^{-5}$ Mpc$^{-3}$), places it in a regime that significantly deviates from the norm for $\Lambda$CDM. If confirmed to lie at $z\approx17$, and if analogues of CEERS-1749 prove to be as common as the first \textit{JWST} extragalactic fields imply, this may provide a compelling constraint on cosmology.
}
\label{fig:UVLF_lambdacdm}
\end{figure*}

\subsection{Implications of the $z\approx17$ scenario}
\label{sec:discussionz17}
\label{sec:lcdm}

If CEERS-1749 is confirmed to lie at $z\approx17$, it would force a major revision of early galaxy evolution models, and potentially even our underlying cosmological framework \citep[see, e.g.,][]{Steinhardt16,Mason22,MBK22}. It is very challenging to produce such extraordinarily luminous and massive galaxies only $\sim$200 Myrs after the Big Bang under standard assumptions in the framework of $\Lambda$CDM cosmology (see also the candidates reported in \citealt{Atek22,Yan22}). 

This situation is demonstrated in Figure \ref{fig:UVLF_lambdacdm}. First, we consider the number density implied by an $M_{\rm{UV}}\approx-22$ source at $z\approx16.5$ found in a search area of a mere 40 arcmin$^{2}$ and $\Delta z=1$. No theoretical UVLF or empirical extrapolation comes close to matching this implied number density within one to several orders of magnitude \citep[][]{Behroozi19,Bowler20,Naidu22,Dayal22}. Even more strikingly, the only way to match such a high number density is by coupling dark matter halo mass functions to a $100\%$ instantaneous star-formation efficiency \citep[see also][]{Mason22}. That is, every baryon allocated to a dark matter halo in accordance with the cosmic baryon fraction is converted into stars immediately following a \citet[][]{Salpeter55} IMF between $0.1-100\,M_{\rm{\odot}}$. For context, the typical efficiency inferred from a variety of arguments at $z\approx6-10$ is $<10\%$ \citep[e.g.,][]{Tacchella18,Stefanon21mass}.

The right panel of Figure \ref{fig:UVLF_lambdacdm} illustrates this extraordinary situation in terms of a stellar mass threshold implied by halo mass functions under a similar assumption of a $100\%$ star-formation efficiency calculated in \citet[][]{Behroozi18}. CEERS-1749 (and its implied number density) places it in a region of the diagram that is in tension with $\Lambda$CDM cosmology. 

Numerous caveats underlie these comparisons, and there are several possible solutions to this apparent tension (CEERS-1749 at $z\sim5$ being the most likely one). Here we detail several other possibilities. 

Lensing is expected to have a substantial effect on the bright end of the UVLF at high redshifts given the increasing optical depth. This is particularly pertinent for CEERS-1749 given that it sits $<2\arcsec$ from a $z\approx5$ $M_{\rm{\star}}\approx10^{11} M_{\rm{\odot}}$ galaxy and that the sightline includes a protocluster. However, lensing corrections even in the most overdense regions tend to be $<2\times$ \citep[e.g.,][]{Mason15lensing}, which does little to alleviate the situation in Figure \ref{fig:UVLF_lambdacdm}. 

Another relevant class of ideas revises the relationship between light and mass. For instance, modifying the IMF to be extremely top-heavy produces much higher UV luminosities for a given stellar mass (up to $\approx10\times$ higher compared to our assumptions of a ``normal" IMF, e.g., \citealt{Fardal07}). Pop III stars and binary stars occurring at low metallicities similarly produce different translations between light and mass. And finally, a possibility that can not be ignored is that some fraction of the luminosity of CEERS-1749 may not be of stellar origin at all, but could arise from accretion onto early black holes \citep[e.g.,][]{Pacucci22}.

\subsection{Implications of the $z\approx5$ scenario}
\label{sec:discussionz5}

\begin{figure*}
\centering
\includegraphics[width=0.9\linewidth]{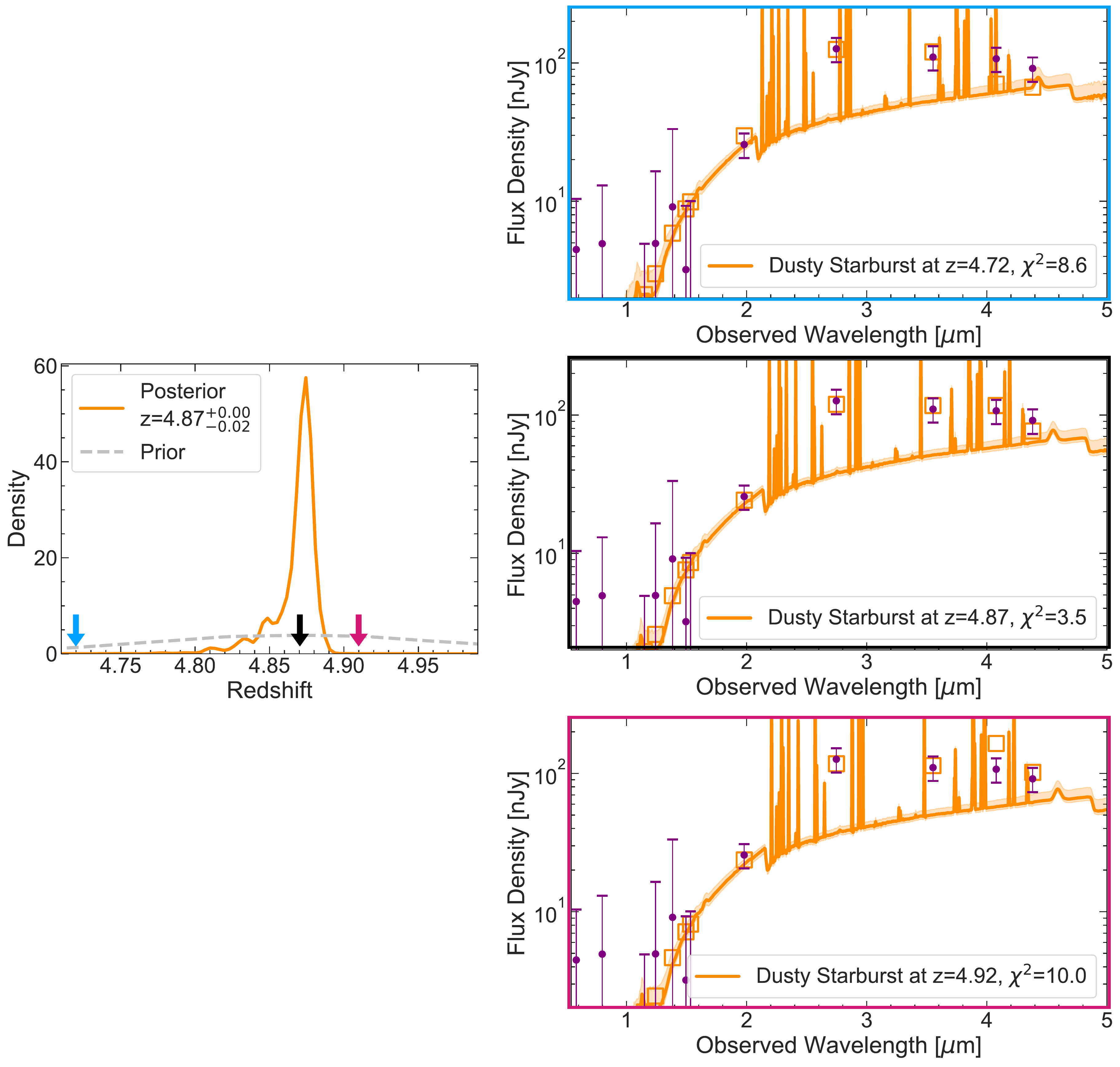}
\caption{Demonstration of the narrow redshift range allowed for a $z\approx5$ interloper with strong emission lines to mimic a $z\approx17$ system. \textbf{Left:} The redshift probability distribution allowed for the dusty star-forming galaxy solution is very narrow because of the stringent requirement to produce a break at $\sim2\mu m$ as well as a blue UV slope across a very wide wavelength range ($\approx2-5\mu$m), which includes a medium band filter (F410M). \textbf{Right:} SEDs of dusty star-forming galaxies with strong emission lines at the most-likely redshift of $z=4.87$ (center, black border), a slightly lower redshift of $z=4.72$ (top, blue border), and a slightly higher redshift of $z=4.92$ (bottom, red border). The medium band at 4.1$\mu$m sensitively discriminates even among these very similar redshifts -- notice the strongly fluctuating $\chi^{2}$ across the panels. This makes an emission line galaxy solution possible only in an very narrow redshift range, severely limiting the volume for such contaminant in $z>10$ searches, particularly when medium bands are used in the filter set.}
\label{fig:conspiracy}
\end{figure*}

We emphasize that the redshift solution for CEERS-1749 across multiple studies, which use diverse data reduction choices and $z>10$ selection techniques, seems unambiguous: $z\approx17$, with $p(z>10)>99.9\%$, and little room permitted for any other possibility \citep[]{Naidu22, Donnan22, Harikane22b}. There is no hint of a $z\approx5$ solution that may be upweighted into relevance by e.g., a luminosity prior. If not for the conservative error floor on the photometry adopted here, and the fortuitous environmental evidence, there would be little reason to place this source at $z\approx5$ \citep[but see also][]{Zavala22}. 

The difficulty of identifying the $z\approx5$ solution for CEERS-1749 could be construed to imply that some fraction of the seemingly secure $z>10$ candidates may be interlopers of the kind discussed in this work. Galaxies with relatively weaker breaks in their SED are the most vulnerable -- e.g., the dusty starburst scenario could account both for their break as well as the slope of their longer wavelength photometry. Such interlopers may help resolve the tension described in the prior section.  At slightly lower redshifts ($z\approx6-10$) the occurrence of the strongest rest-optical lines as well as the presence of both Balmer breaks as well as Lyman breaks in the NIRCam coverage provide additional safeguards \citep[e.g.,][]{Labbe22}. Further, MIRI photometry (e.g., see how the dusty galaxy stands out in Table \ref{table:miri}), an additional medium band (for example, in the JADES GTO program filter-set, \citealt{Rieke20JADES}), or any spectroscopy would comfortably protect against such interlopers.

We also emphasize that we are dealing with an extraordinary situation given the foreground protocluster. The redshift range in which strong emission lines in a dusty system perfectly conspire to mimic a Lyman break \textit{as well as} the blue UV-slope of a $z>10$ galaxy is very narrow. For instance, in our dusty starburst scenario the full redshift posterior collapses to an extremely tight $z=4.87^{+0.00}_{-0.02}$. The implied foreground contaminant volume for such a narrow redshift range is therefore far less dire for $z>10$ searches than may seem at first glance, particularly when a medium-band is included in the filter-set. We illustrate this situation in Figure \ref{fig:conspiracy}.

If this source turns out to be a quiescent galaxy, it would be the highest redshift quiescent system known, with the current most distant galaxy at $z\approx4$ \citep[][]{Valentino20}. Finding quenched galaxies at such early times places stringent constraints on galaxy evolution models and the physics of feedback. Since the galaxy has a relatively low stellar mass, it is expected to be undergoing ``environmental quenching" \citep[e.g.,][]{Peng10}, which is consistent with its location in an overdensity. However, it being at such high redshift, only $\approx1$ Gyr after the Big Bang, challenges many theoretical scenarios that have been developed at lower redshifts, including the general result that satellites experience ``delayed-then-rapid" quenching \citep[e.g.,][]{Wetzel13,Fillingham19,Naidu22MZR}, which starts several Gyrs after infall.

On the other hand, confirmation of the dusty starburst scenario would extend the \textit{HST}-dark population to lower masses, raising intriguing questions about how such low-mass galaxies got so dusty so fast in only the first billion years of the Universe \citep[e.g.,][]{Ferrara16,Popping17,Lesniewska19, Dayal22}. Further, protoclusters at high redshift are expected to be among the first sites of star-formation and reionization in the Universe -- the dozens of potentially associated neighbors around CEERS-1749 are therefore exciting targets that make multi-object spectroscopic follow-up an even more compelling proposition.

\section{Summary \& Outlook}
\label{sec:summary}

Within the first few weeks of \textit{JWST}'s initial data release, the facility has already delivered a major expansion of our cosmic frontier, with dozens of $z>10$ candidates being reported. Several of these sources are unexpectedly luminous ($M_{\rm{UV}}\lesssim-21$), and are far more common than state-of-the-art projections for the \textit{full} Cycle 1 yield across all scheduled programs. In this paper we present potentially the most extreme of these systems: CEERS-1749. We discuss two redshift solutions for the source, each of which has wide-ranging implications.

\begin{itemize}
  
    \item Across a variety of SED-fitting choices, we find $z\approx17$ to be the most likely redshift with no lower-$z$ solutions found. Other independent analyses and state-of-the-art techniques find a similarly confident, unambiguous solution \citep[e.g.,][]{Donnan22,Harikane22b}.  Only a conservative $\approx20\%$ error-floor on all photometry to account for systematic uncertainties shows hints of a $z\approx5$ solution. [Figure \ref{fig:summaryCEERSz17}]  
    \item If the galaxy is at $z\approx17$, then it has physical properties (e.g., $\beta_{\rm{UV}}$, SFR) and a morphology (clumpy in the rest-UV) expected of $z>10$ galaxies. However, most strikingly, its stellar mass ($\approx5\times10^{9} M_{\rm{\odot}}$) and UV luminosity ($M_{\rm{UV}}\approx-22$) are unexpected for a system lying a mere $\sim220$ Myrs from the Big Bang. [Table \ref{table:properties}, Figure \ref{fig:UVLF_lambdacdm}]

    \item We show that the SED of the galaxy can be explained by a $\approx10^{9}-10^{10} M_{\rm{\odot}}$ quiescent galaxy with line emission arising from ionized gas, or a $\approx5\times10^{8} M_{\rm{\odot}}$ dusty starburst whose nebular emission lines boost the photometry and conspire to produce an apparently blue slope in the $>2\mu$m photometry. The morphology of the source, which shows hints of a merger and/or tidal disturbance supports both these scenarios. [\S\ref{sec:whatisz5}, Figure \ref{fig:altSEDs}, \ref{fig:sizefits}]
    
    \item The $z\approx5$ scenario has strong environmental support. The three nearest neighbors of the galaxy lie at precisely the redshift required by the low-$z$ solution, i.e., $z\approx5$. Further, there is a hint of a substantial $z\approx5$ overdensity in the CEERS field -- $\approx25$ sources $<5'$ from the candidate have photometric redshifts of $z\approx5$, some of them displaying mature stellar populations expected of an early-forming protocluster. [\S\ref{sec:lowz}, Figure \ref{fig:z5solution}]
    
    \item These results suggest that at certain specific redshifts, $z>10$ candidates from \textit{JWST} may contain a class of lower-$z$ interlopers.  If not for the various lines of circumstantial evidence, there would have been little reason to doubt the $z\approx17$ solution. However, we stress that the perfect storm of parameters required to mimic both the break and continuum of a $z>10$ candidate is possible only in a very narrow redshift range, especially when medium-bands are employed, implying that such interlopers may not be a major concern for $z>10$ searches. [\S\ref{sec:discussionz5}, Fig. \ref{fig:conspiracy}]
    
\end{itemize}

Spectroscopic follow-up of this remarkable galaxy is of critical urgency to \textit{JWST}'s mission of expanding the cosmic frontier. A $z\approx5$ solution might provide new insights into the physics of quenching and dust production, as well as an important class of interloper galaxies to strengthen $z>10$ searches. On the other hand, if this source does lie at $z\approx17$, we may embark on the grand enterprise of revising the physics of galaxy evolution at the earliest epochs.

\facilities{\textit{JWST}, \textit{HST}}

\software{
    \package{IPython} \citep{ipython},
    \package{matplotlib} \citep{matplotlib},
    \package{numpy} \citep{numpy},
    \package{scipy} \citep{scipy},
    \package{jupyter} \citep{jupyter},
    \package{Astropy}
    \citep{astropy1, astropy2},
    \package{grizli}
    \citep{grizli}
    }
    
\acknowledgments{
We are grateful to the CEERS team for planning these early release observations and speedily making resources available to the community that have made this work possible. 

RPN acknowledges funding from \textit{JWST} programs GO-1933 and GO-2279. We acknowledge support from: the Swiss National Science Foundation through project grant 200020\_207349 (PAO, LB, AW).
The Cosmic Dawn Center (DAWN) is funded by the Danish National Research Foundation under grant No.\ 140.
RJB and MS acknowledge support from NWO grant TOP1.16.057. S. Bose is supported by the UK Research and Innovation (UKRI) Future Leaders Fellowship [grant number MR/V023381/1]. MS acknowledges support from the CIDEGENT/2021/059 grant, and from project PID2019-109592GB-I00/AEI/10.13039/501100011033 from the Spanish Ministerio de Ciencia e Innovaci\'on - Agencia Estatal de Investigaci\'on. S. Belli is supported by the Italian Ministry for Universities and Research through the \emph{Rita Levi Montalcini} program. K.E.H. acknowledges support from the Carlsberg Foundation Reintegration Fellowship Grant CF21-0103. PD acknowledges support from the NWO grant 016.VIDI.189.162 (``ODIN") and from the European Commission's and University of Groningen's CO-FUND Rosalind Franklin program. 

Cloud-based data processing and file storage for this work is provided by the AWS Cloud Credits for Research program.

This work is based on observations made with the NASA/ESA/CSA James Webb Space Telescope. The data were obtained from the Mikulski Archive for Space Telescopes at the Space Telescope Science Institute, which is operated by the Association of Universities for Research in Astronomy, Inc., under NASA contract NAS 5-03127 for \textit{JWST}. These observations are associated with programs \# 1324 and \# 1345.
}

\bibliography{MasterBiblio}
\bibliographystyle{apj}

\end{CJK*}
\end{document}